\documentclass[preprint,aps]{revtex4}
\usepackage{epsfig}
\usepackage{bm}
\begin{document}

\title
{\bf Strongly compressible current sheets under gravitation}
\author{\normalsize{S. I. Vainshtein$^1$, Z. Miki\'c$^2$, F. Cattaneo$^{1,3}$,
R. Rosner$^{1,3}$, and R. Sagdeev$^4$}}
\affiliation{\small \centerline{\it $^1$University
of Chicago}
\centerline{\it $^2$Science Applications International Corporation (SAIC)}
\centerline{\it $^3$Argonne National Laboratory}
\centerline{\it $^4$University of Maryland}}

\begin{abstract}
{Many stormy events in astrophysics occur due to the sudden magnetic energy
release. This is possible if a magnetic configuration abruptly changes its topology,
an event usually referred to as magnetic reconnection. It is  known that pure
Ohmic decay is inefficient, occurring during cosmological times
(due to the huge characteristic scales $L$). It is recognized
that the presence of current sheets speeds up the process, but still
insufficiently$^{1,2,3,4,5}$.
We show that, in highly compressible and substantially gravitational media, the
reconnection is fast enough to account for stormy events. Thus,
highly compressible situations offer exiting opportunities in
explanations of violent events, although full-scale compressible
and gravitational simulations proved to be quite challenging.}
\end{abstract}
\maketitle

Basically, there are two characteristic times: Huge Ohmic decay time, $t_o=L^2/\eta$, 
($\eta$ is resistivity), and
sufficiently short Alfv\'en time, 
$$
t_A=\frac{L}{C_A},
$$
$C_A$ is Alfv\'en velocity, 
$
C_A={B}/{\sqrt{4\pi\rho}},
$
the ratio of these times being the Lundquist number $S\gg1$. The Alfv\'en time would be
good enough to explain many stormy events in astrophysics. The problem is that the magnetic
field would not change its topology that fast. In particular,  the Sweet-Parker (SP)  
current sheet$^{1,2}$, with the width
$$
\delta_{SP}=\frac{L}{S^{1/2}}.
$$
results in shorter  (than Ohmic) reconnection time,
$
t_{SP}=t_o/S^{1/2}=t_AS^{1/2}.$
 The (dimensionless) reconnection rate for SP is
\begin{equation} 
R=\frac{t_A}{t_{SP}}=\frac{1}{S^{1/2}}.
\label{rate}
\end{equation}
For huge astrophysical values of  $S=10^{10\div20}$, say, the SP mechanism is too slow. 
In other words, the exponent $1/2$ in (\ref{rate}) is too big.

Much more efficient mechanism was suggested by Petschek$^3$, with 
reconnection rate,
$$
R=\frac{t_A}{t_P}=\frac{\pi}{8\ln{S}},
$$
which is sufficiently fast.
 It was pointed out however that this mechanism may only work under special conditions
in the vicinity of the x-point, where the reconnection occurs$^{4,5,6,7}$, in 
particular, nonuniform resistivity$^{8,9}$. 
 The aim of this 
letter is to argue and present numerical evidence that strong compressibility and gravity  result in such
conditions, leading to Petschek-like reconnection.

\begin{figure}
\psfig{file=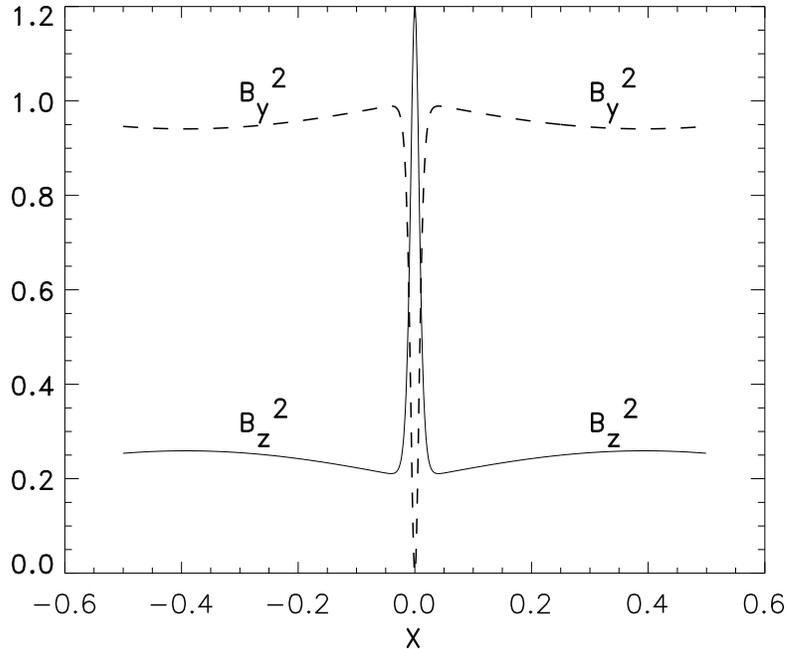}
\caption{Typical magnetic distribution in the vicinity of the current sheet. $B_y$-component
abruptly changes sign at current sheet location $x=0$, and therefore $B_z$-component acquires a sharp maximum.}
\label{fig1}
\end{figure}

We will deal with 2.5-dimensional case, when magnetic field is presented as
${\bf B}=\{{\bf B}_\perp,B_z\}=\{B_x(x,y),B_y(x,y),B_z(x,y)\}$.
In the vicinity of the current sheet the configuration is in nearly 1-dimensional equilibrium, i.e.,
\begin{equation}
p+\frac{B_y^2+B_z^2}{8\pi}={\rm const}, 
\label{equilibrium}
\end{equation}
where $p$ is the pressure. In low pressure plasma (of stellar coronas), $\beta=p/(B^2/8\pi)\ll 1$,
we can neglect the pressure in (\ref{equilibrium}), leading to a field configuration
depicted in Fig.
\ref{fig1}: With sharp maximum of the $B_z$-component at $x=0$. This makes $x=0$ point 
singular$^{10}$ because this maximum will be "wiped out" by Ohmic dissipation. Indeed,
it takes only $\delta_{PS}^2/\eta=t_A$, i.e., Alfv\'en time, for this maximum to disappear.
As a result, the current sheet collapses, leading to density and pressure build-up, and
speeding-up the reconnection rate. More detailed estimates$^{11}$ indeed show that, in the
vicinity of the current sheet $\nabla\cdot{\bf v}<0$, i.e., compression, and $\nabla\cdot{\bf v}
\approx 1/t_A$. 
Thus, substantial compressibility creates a singularity -- i.e., special conditions in the 
vicinity of the current sheet.

Still, the pressure build-up would try to stifle the collapse, and presumably slow-down the
reconnection. We show that the gravity, present in astrophysical conditions
 creates
additional special conditions and again facilitates the reconnection. If the scale
height $H$ is $< L$ (quite a modest requirement, satisfied in the Sun) the equilibrium 
condition (\ref{equilibrium}) is incompatible with the
gravity, unless the current sheet is strictly horizontal, which is unlikely to happen.
In the vicinity of the current sheet, the magnetic field is a function of $x$ 
only, while
gravity imposes $\sim\exp(-y/H)$-dependence for $p$, and (\ref{equilibrium}) cannot be satisfied. 

Thus, the reconnection starts to ``feel" the gravitation when $L/H>1$. In addition,  we note, that 
dramatic
build-up of density and therefore of pressure perturbations,  is unlikely in the
stellar atmospheres. Indeed, unless there are  special conditions (satisfied, e.g., for the
 prominences), in the presence of gravity forces, the density ``blobs" cannot be sustained
in coronas for a long time, and they would slip down along the magnetic field lines.

To be more specific, for strongly gravitational stars, when more strict inequality, 
$L/H>\beta^{-1/2}$ is satisfied, or
$$
Z=\frac{L\beta^{1/2}}{H}=\frac{Lg}{C_A^2\beta^{1/2}}> 1,
$$
the fall down time is less than Alfv\'en time.
We expect that, if $Z\gg 1$, then, roughly
speaking, the matter remains in stratified and almost unperturbed state all the time during the
reconnection process. This prevents
the build-up of the pressure in the vicinity of the current sheet. Simple but more
cumbersome estimates$^{11}$ 
 confirm this statement.

\begin{figure}
\psfig{file=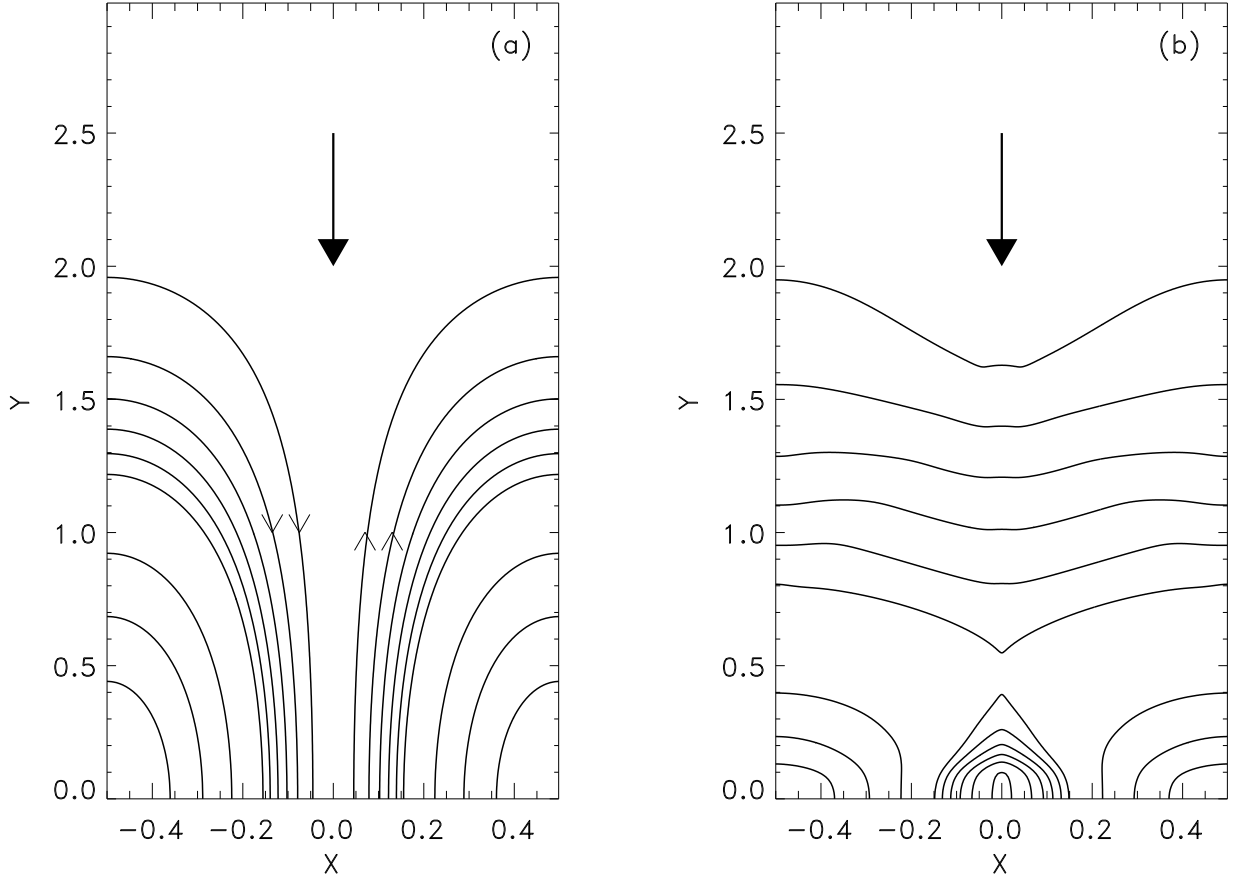}
\caption{Initial (a) and final (b) magnetic configuration in our simulations. Big
vertical arrow represents the gravity force.}
\label{fig1a}
\end{figure}

Of course, the only reliable way to check these statements is to use numerical simulations.
There are not so many compressible simulations although it is known that the
compressibility would speed up reconnection process$^{12}$, and that was
indeed observed experimentally$^{13}$. The compressibility has a profound effect
in the vicinity of magnetic null-points$^{14}$. 

We provide ``global" simulations,
i.e., for the whole region. This makes it possibly
to study the scaling exponents of the reconnection rate.
We present three
types of simulations:
\begin{figure}
\psfig{file=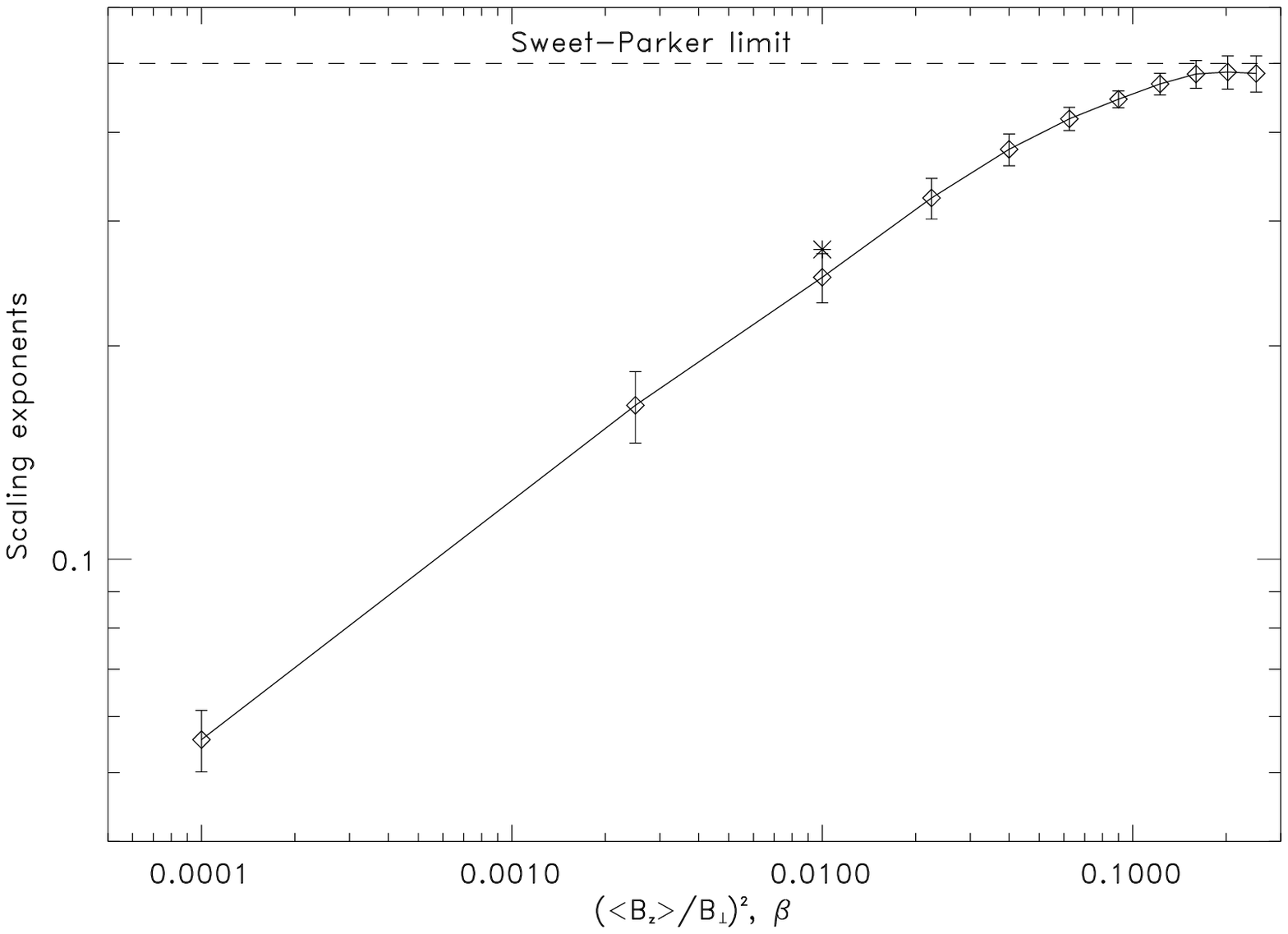}
\caption{The scaling exponents depending on compressibility. The diamonds  correspond
to the different values of $(B_z/B_\perp)^2$, and $Z\gg 1$, while the asterisk 
corresponds  simulations with  $\beta=0.01$ and $Z=0$.}
\label{fig12}
\end{figure}

\begin{figure}
\psfig{file=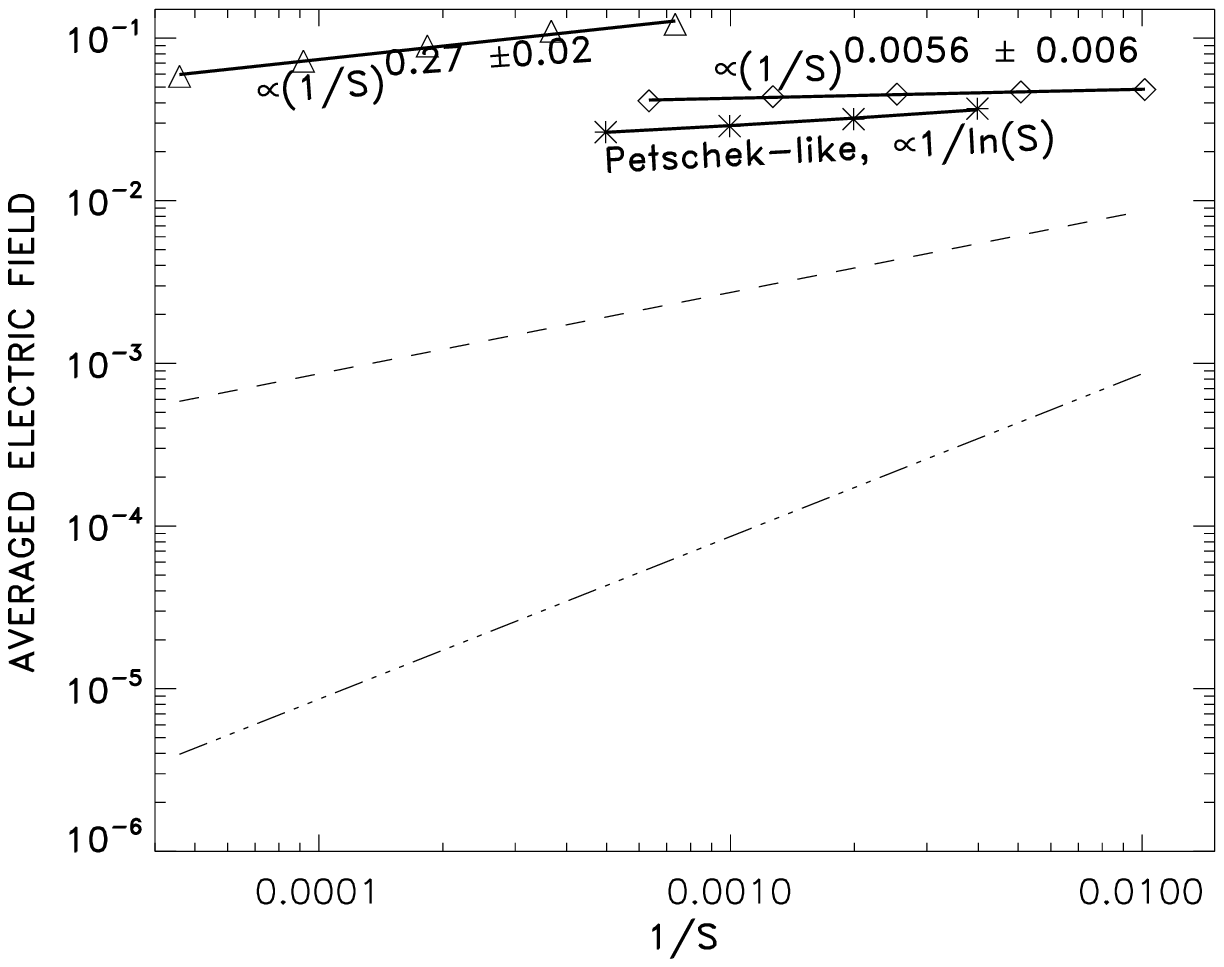}
\caption{Normalized averaged electric field measured in simulations.
Triangles correspond to $Z=0$ simulations of rosette-structure. Diamonds -- to $Z\gg1$, highly
compressible case.
Asterisks correspond to $Z=0.2$, solar conditions, fitted with Petschek-like scaling.
 Short dashes correspond to Sweet-Parker, $\sim 1/S^{1/2}$, and the
dashed-dotted line corresponds to pure Ohmic decay, $\sim 1/S$.}
\label{fig4}
\end{figure}

\noindent
1) $Z=0$, no gravity, but substantial compression. A rosette-structure configuration was considered, 
in which case the
current sheet formation is inevitable$^{15}$. 
The plasma collapse was indeed persistently observed in simulations$^{16,11}$.
Thus, it was estimated that in order to reach equilibrium Fig. \ref{fig1}, $B_z$ should
increase 1.5 times, say, resulting in pressure increase only twice, while the
observed pressure was increasing up to 30 times! 

This process strongly depends
on the compressibility: With increasing $B_z$-component, when plasma becomes nearly incompressible, old 
results about SP slow reconnection rate$^{4,5,6,8}$. 
Note, however, that rosette-structure simulations cannot reach very high compressibility: If we take
very small initial $B_z$-component, the magnetic configuration will collapse in such a way that $B_z$ is
increased. Nevertheless, they definitely show that the current sheet collapse speeds-up the reconnection
process: The reconnection rate scaling exponent is $=0.27\pm 0.02$, definitely smaller than $1/2$, characteristic 
for the SP process.

\noindent
2) $Z\gg 1$, extremely strong gravity. 
We can imagine that two pairs of ``sunspots" of opposite polarities are initially far away from
each other, and therefore they are not connected. Suppose that, in the course of
evolution, they approach each other. Due to frozen-in conditions, the topology of the field
lines is not easy to change, and, for a while, they remain disconnected from each
other, Fig. \ref{fig1a}(a). Due to finite conductivity, a current sheet will form, 
and the reconnection
starts$^{17,18}$. Further evolution is observed in our numerical simulations, with
final (reconnected) state depicted in Fig. \ref{fig1a}(b).

This time, the $B_z$-component can be arbitrary small, and therefore we can study extreme compressibility.
We measured electric field (responsible for high energy particles acceleration)
 normalized  on a maximal
possible field, $\bf v\times B$, where $v=C_A$, (which is essentially the
reconnection rate, like (\ref{rate})), for different Lundquist numbers $S$, to get the 
scaling exponents. We also provided simulations with different compressibility's, characterized by parameter
$B_z/B_\perp$. 
When this parameter is small,  the media is highly compressible, while increasing
this parameter, we will approach essentially incompressible situation. As seen from
Fig. \ref{fig12}, the incompressible media is reached when $(B_z/B_\perp)^2=.04$, and
then the Sweet-Parker reconnection is recovered. For higher compressibility's we found
that the reconnection is much faster (i.e., the scaling exponents are mach smaller). The fastest reconnection proceeds essentially with
Alfv\'en time.
It is obvious that the compressibility speeds up the reconnection process in the presence
of strong gravitation.

\noindent
3) Full-scale calculations with arbitrary $Z$, and arbitrary compressibility's are not easy
to perform. We managed to simulate conditions for the Sun: $L/H>1$, so that special conditions
do appear at the vicinity of the x-point due to the gravity, although $Z=0.2\div0.3$. So far,
we fail to simulate typical $\beta=0.01$, or so, but we do have the case $\beta=0.001$ 
(the field is 3 times stronger than typical). 
The results for all three types of simulations
are summarized in Fig. \ref{fig4}. For solar conditions, the scaling is close to 
Petschek-like rate, the standard deviation of measured
electric field from Petschek-like being $=0.01$.  The simulations reveal rather
stormy developments: Creation of a current sheet, and fast reconnection, see
Supplement movie$^{19}$.

All these simulations show the importance of strong compressibility and gravity forces
in explaining stormy and violent magnetic events in astrophysics.

\acknowledgments{We thank E.N. Parker and  B.C. Low
for numerous 
discussions,  and A. Obabko for helping with simulations.
This work was supported by the NSF sponsored Center for Magnetic
Self-Organization at the University of Chicago and NASA grant  NNG04GD90G.}

\newpage

\centerline{References}

\noindent
1.
Sweet, P.A.  The neutral point theory of solar flares.
{\it Electromagnetic Phenomena in Cosmical Physics}, edited by B. 
Lehnert (Cambridge Press, New York, 1958), p. 123.

\noindent
2.
Parker, E.N. Sweet's mechanism for merging magnetic fields in conducting
fluids.
{\it J. Geophys. Res.} {\bf 62}, 509 (1957).

\noindent
3.
Petschek, H.E. Magnetic field annihilation.
{\it AAS-NASA Symposium of the Physics of Solar Flares},
NASA-SP 50, edited W.N. Ness (National Aeronautics and Space Administration,
Washington, DC 1964), p. 425.

\noindent
4.
Biskamp, D. Magnetic reconnection via current sheets.
{\it Phys. Fluids} {\bf 29}, 1520 (1986).

\noindent
5.
Biskamp, D. {\it Magnetic Reconnection in Plasmas} (Cambridge University Press, New
York, 2000), p. 239

\noindent     
6. Priest, E.R. \& Forbes, T.G. {\it Magnetic Reconnection} (Cambridge 
University Press, Cambridge, 2000) .

\noindent
7.
Jemella, B.D., Shay, M.A \& Drake, J.F. \& Rogers, B.N. Impact of Frustrated
Singularities on Magnetic Island Growth.
 {\it Phys. Rev. Letters} {\bf 91}, 125002 
(2003).

\noindent
8.
Biskamp D. \& Schwarz, E. Localization, the clue to fast magnetic reconnection.
{\it Physics of Plasmas}, {\bf 8}, 4729 (2001)

\noindent  
9. Baty, H. Priest, E.R. \& Forbes, T.G. Effect of nonuniform resistivity 
in Petschek reconnection.
{\it Physics of Plasmas}, {\bf 13}, 022312 (2006)

\noindent
10.
Low, B.C. Nonlinear classical diffusion in a contained plasma.
{\it Phys. Fluids} {\bf 25}, 402 (1982).

\noindent
11.
Vainshtein, S.I., Miki\'c, Z. \& Sagdeev, R.Z. ``Compression of the current sheet and 
its impact into the reconnection rate", arxiv.org/abs/0711.1666v1 (2007).

\noindent
12.
Parker, E.N. The Solar-Flare Phenomenon and the Theory of Reconnection 
and Annihilation of Magnetic Fields.
 {\it Astrophys. J. Supplement}, {\bf 8}, 177 (1963)

\noindent
13.
Ji, H., Yamada, Hsu, M.S., Kulsrud, R., Carter, T. \& Zaharia, S. 
Magnetic reconnection with Sweet-Parker characteristics 
in two-dimensional laboratory plasmas. {\it Phys. Plasmas},
{\bf 6}, 1743 (1999).

\noindent 
14. Pontin, D.I., Bhattacharjee, A.\&  Galsgaard, K. 
Current sheets at three-dimensional magnetic nulls: 
Effect of compressibility. {\it Phys. Plasmas}, {\bf 14},
052109 (2007)

\noindent
15.
Vainshtein, S.I., Miki\'c, Z., Rosner, R. \& Linker, J.
Evidence for topological nonequilibrium in magnetic
configurations.
{\it Phys. Rev. E}, 
{\bf 62}, 1245 (2000).

\noindent
16.
Miki\'c, Z., Vainshtein, S.I., Linker, J. \& Rosner, R. in preparation.

\noindent
17.
Syrovatsky, S.I. Formation of currentsheets in a plasma with a frozen-in strong
magnetic field.
{\it Sov. Phys. JETP}, {\bf 33}, 933 (1971).

\noindent
18.
Low, B.C. 
On the possibility of electric-current sheets in dense formation.
{\it Physics of Plasmas}, {\bf 14}, 122904 (2007).

\noindent
19. Watch the movie on the Web,
$\rm http://flash.uchicago.edu/\tilde{}~samuel/Gravitational\_Reconnection$.

\end{document}